\documentclass[psfig]{mn2e} 
\usepackage{psfig}

\def\sn{\hbox{S/N}}

\def\vsin{\hbox{$v \sin i$}}  
  
\def\kms{\hbox{km\,s$^{-1}$}}  
\def\ms{\hbox{m\,s$^{-1}$}}

\def\em{\it}  
  
\def\degr{\hbox{$^\circ$}}

\def\kis{\hbox{$\chi^2$}}   
\def\kisr{\hbox{$\chi^2_{\rm r}$}}

\def\xib{\hbox{$\xi$~Bootis~A}}

\begin{document}  

\title[Magnetic field of the G8 dwarf $\xi$~Bootis~A]  
{Large-scale magnetic field of the G8 dwarf $\xi$~Bootis~A}
  
\makeatletter  
  
\def\newauthor{%  
  \end{author@tabular}\par  
  \begin{author@tabular}[t]{@{}l@{}}}  
\makeatother  
   
\author[P.~Petit et al.]  
{\vspace{1.5mm}
P.~Petit$^{1}$, J.-F.~Donati$^2$, M.~Auri\`ere$^2$, J.D.~Landstreet$^3$, F. Ligni\`eres$^2$\\
{\hspace{-1mm}\vspace{1.5mm}\LARGE\rm S.~Marsden$^{4}$, D. Mouillet$^2$, F. Paletou$^2$, N.~Toqu\'e$^2$, G.A.~Wade$^5$}\\
$^1$Max-Planck Institut f\"ur Sonnensystemforschung, Max-Planck-Str. 2, 37191 Katlenburg-Lindau, Germany ({\tt petit@linmpi.mpg.de})\\
$^2$Laboratoire d'Astrophysique de Toulouse-Tarbes, Observatoire Midi-Pyr\'en\'ees, 14 Av.\ E.~Belin, F--31400 Toulouse, France\\ ({\tt donati@ast.obs-mip.fr, auriere@ast.obs-mip.fr, francois.lignieres@obs-mip.fr, fpaletou@ast.obs-mip.fr}\\
{\tt mouillet@bagn.obs-mip.fr, toque@ast.obs-mip.fr})\\
$^3$Department of Physics and Astronomy, The University of Western Ontario, London, Ontario, Canada, N6G 3K7\\ ({\tt jlandstr@astro.uwo.ca})\\
$^4$Institute of Astronomy, ETH Zentrum, CH-8092 Z\"urich, Switzerland ({\tt marsden@usq.edu.au})\\
$^5$ Royal Military College of Canada, Department of Physics, P.O. Box 17000, Station "Forces", Kingston, Ontario, Canada, K7K 4B4\\ ({\tt Gregg.Wade@rmc.ca })\\
}

\date{2005, MNRAS}  
\maketitle  
   
\begin{abstract}   
We investigate the magnetic geometry of the active G8 dwarf $\xi$~Bootis~A, from spectropolarimetric observations obtained in 2003 with the MuSiCoS \'echelle spectropolarimeter at the T\'elescope Bernard Lyot (Observatoire du Pic du Midi, France). We repeatedly detect a photospheric magnetic field, with periodic variations consistent with rotational modulation. Circularly polarized (Stokes V) line profiles present a systematic asymmetry, showing up as an excess in amplitude and area of the blue lobe of the profiles. A direct modeling of Stokes V profiles suggests that the global magnetic field is composed of two main components, with an inclined dipole and a large-scale toroidal field. We derive a dipole intensity of about 40~G, with an inclination of 35\degr\ of the dipole with respect to the rotation axis. The toroidal field strength is of order of 120~G. A noticeable evolution of the field geometry is observed over the 40 nights of our observing window and results in an increase of the field strength and of the dipole inclination.  

This study is the first step of a long-term monitoring of $\xi$~Bootis~A and other active solar-type stars, with the aim to investigate secular fluctuations of stellar magnetic geometries induced by activity cycles.   
\end{abstract}  
  
     
\section{Introduction}   
\label{sect:introduction}  

Fluctuations in the magnetic activity of the Sun are dominated by the well-known solar cycle, historically discovered as a $\approx$11~yr period of the sunspot number (Schwabe 1843), which itself was later linked to a polarity reversal in bipolar magnetic regions between both hemispheres of the Sun. The global activity cycle of the Sun, involving two consecutive polarity switches, is therefore of period $\approx$22~yr (Hale \& Nicholson 1925). Efforts to develop a theoretical explanation of such a cyclical activity have given rise to the so-called solar dynamo theories (see e.g. Ossendrijver 2003 for a review).

In solar-type stars, long-term monitoring of chromospheric emission suggests that the existence of activity cycles depends a lot on the evolutionary stage and rotation rate (Baliunas et al. 1995). It is suggested in particular that a cyclical activity is common among slowly-rotating, low activity main sequence stars similar to the Sun. By contrast, younger stars (rotating faster and possessing a much higher level of activity) generally display more erratic fluctuations of activity level. If one is to understand the dynamo processes at work in solar-type stars, and how various stellar parameters may affect the onset of different types of dynamos, it is necessary to obtain direct information about the geometry of stellar magnetic fields and about secular fluctuations of this geometry. 

For rapidly rotating stars, the surface distribution of the magnetic field vector is now routinely reconstructed with the technique of Zeeman-Doppler Imaging (ZDI hereafter, Donati \& Brown 1997, Donati et al. 2003). Mapping techniques are however much more limited for slow rotators, because in their case all magnetic regions of the visible hemisphere of the star produce Zeeman signatures of similar radial velocities. Because of the usually complex field topology of cool active stars, spectroscopically superimposed Zeeman signatures of mixed polarities are observed at any time, resulting in an important cancellation of the polarized signal. This situation is occurring for slow rotators of course, but also for low-mass stars with small radii and for stars with low inclination angles. For such objects, most individual active regions remain unresolved. The only measurable quantity is the global field of the star, i.e. the magnetic field that remains after integration over the visible stellar hemisphere. 

In the case of the Sun, the large-scale structure of the photospheric field can be determined using magnetograph data (e.g. Stenflo 1991). Outside the belts of active regions, and in particular around the poles, a large-scale component of the surface field is observed (sometimes called the {\em background field} of the Sun). This component was first investigated by Babcock \& Babcock (1955), who report that its geometry is mostly dipolar and close to axi-symmetry. The solar background field is modulated by the Hale cycle, in anti-phase to the sun-spot cycle (its polarity  reversal occurring at solar maximum). The strength of the field at the pole during solar minimum is of the order of 5~G (Smith \& Balogh 1995). Since its cyclical evolution is controlled by the large-scale dynamo of the Sun, investigating the geometry and secular evolution of global fields on a sample of slowly-rotating late-type stars may also bring precious information about the dynamo processes at the root of their magnetic activity. 

As a first attempt in this direction, we concentrate on the solar-type star \xib. From photospheric and chromospheric activity tracers, this object can be considered as a very active star, with irregular fluctuations of activity (Toner \& Gray 1988, Baliunas et al. 1995, Gray et al. 1996). Its surface magnetic field was first detected by Robinson et al. (1980). The rotational modulation of the large-scale field was then investigated by Plachinda \& Tarasova (2000) from a time series of longitudinal field measurements collected over some two decades. We propose here a different and potentially more powerful approach, making use of a time-series of high signal-to-noise spectropolarimetric data, and based on a direct modeling of the polarized line profiles, from which we derive tight constraints on the field geometry. 

The article is divided as follows. We first discuss the fundamental parameters of \xib. We then present our set of spectropolarimetric data and investigate the large-scale geometry of the photospheric field. We discuss our results in relation with previous works on magnetically active stars and propose some directions in which this study may be extended in a near future.

\section{Fundamental parameters}
\label{sect:param}

$\xi$ Bootis A is a bright member ($m_v$ = 4.7) of one of the nearest visual binary systems (parallax $\pi = 0.1491$ translating into a distance of 7~pc, Gliese \& Jahreiss 1991). The system has been regularly observed for decades and is now known to follow a 151~yr orbital period. The masses of $\xi$~Boo~A and B are estimated to be 0.85 and 0.72 $M_\odot$ respectively and the inclination angle of the orbital plane is 140\degr\ (Wielen 1962). The respective spectral types of both components is G8V and K5V (Abt 1981). The mean effective temperature of the primary component ($T_{\rm eff}\approx 5550$~K) was determined by Gray (1994), using line-depth ratios corrected from metalicity dependence. Noticeable rotational and secular fluctuations of $T_{\rm eff}$, presumably produced by magnetic activity, are also reported by Gray et al. (1996) and reach a level of 12~K. The metalicity of the primary dwarf is $[{\rm Fe/H}]=-0.20\pm0.08$ (Cayrel de Strobel et al. 1992) and its luminosity is $Log(L/L_\odot)=-0.26\pm0.03$ (Fernandes et al. 1998). Using stellar evolutionary models for both stars of the system, Fernandes et al. (1998) derive different additional stellar parameters. They propose an helium abundance $Y=0.26\pm0.02$ and a mixing length parameter $\alpha_{MLT}=1.7\pm0.03$. Their models are consistent with a young age ($2\pm2$~Gyr), which is also suggested by the high chromospheric emission (Baliunas et al. 1995) and high lithium abundance (Herbig 1965).

The rotation period has been estimated in several studies, making use of different rotationally modulated tracers (Ca II emission by Noyes et al. 1984, line asymmetry and line ratios by Toner \& Gray 1988, longitudinal magnetic field by Plachinda \& Tarasova 2000). All periods are in the 6.1-6.5~d range, but the tight error bars on each individual estimate reveal an apparent discrepancy. The shortest period is derived from the integrated longitudinal magnetic field (P$\approx$6.1455~d). The longest one is obtained from variations in line bisectors (P$\approx$6.43~d). We use this last period, carefully derived by Toner \& Gray (1988) from a high-quality data set, as our reference to determine the rotational phases of our observations. We will show in Sect. \ref{sect:beff} that this estimate is also consistent with the rotation period derived from our own observations. We take Julian date 2,452,817.41 (the date of our first observation) as phase zero. 

From the luminosity estimated by Fernandes et al. (1998), we deduce a stellar radius $R=0.8\pm0.03~R_\odot$. Combining this value with the projected rotational velocity \vsin$=3\pm0.4$~\kms\ (Gray 1984), we obtain an inclination angle $i=28\pm5$\degr, which gives also the equatorial velocity $v_e=6.4\pm0.3$~\kms. In previous studies (Toner \& Gray 1988 and after them Plachinda \& Tarasova 2000), the rotation axis was assumed to be perpendicular to the orbital plane of the binary system, i.e. $i$ was chosen close to 40\degr. However, considering the very long orbital period of the system, the hypothesis of a tidally-induced perpendicularity of the rotation axis with respect to the orbital plane is not expected (Zahn 1977). We therefore prefer our own estimate of $i$, which is adopted in the rest of the paper.  

\section{Instrumental setup and observations}
\label{sect:obs}

\begin{table*}
\caption[]{Journal of observations. From left to right, we list the date of observations, the Julian date, the UT time, the \sn\ of raw Stokes V spectra and of LSD mean profiles, the rotation cycle (according to a 6.43~d period and taking our first observation as phase zero), the longitudinal field $B_l$ and the uncertainty $\sigma_B$ on the longitudinal field measurement.}
\begin{tabular}{cccccccccccc}
\hline
Date  & JD           & UT & \sn  & \sn   & rotation cycle   & $B_l$  &  $\sigma_B$\\
      & (+2,450,000) & (hh$:$mm$:$ss)   & V    & V$_{\rm LSD}$ & & (G) & (G)\\
\hline
2003 Jun 26 & 2817.41 & 21:52:47 & 560 & 19721 & 0.0000 & 4 & 3 \\
2003 Jun 27 & 2818.39 & 21:15:10 & 440 & 13146 & 0.1514 & -2 & 5 \\
2003 Jun 28 & 2819.37 & 20:56:31 & 500 & 17103 & 0.3050 & 0 & 4 \\
2003 Jul 04 & 2825.38 & 21:02:31 & 540 & 18973 & 1.2387 & 4 & 4 \\
2003 Jul 05 & 2826.38 & 21:05:42 & 490 & 16411 & 1.3946 & 14 & 5 \\
2003 Jul 06 & 2827.37 & 20:56:51 & 510 & 16912 & 1.5492 & 15 & 4 \\
2003 Jul 07 & 2828.39 & 21:19:35 & 430 & 14239 & 1.7071 & -2 & 5 \\
2003 Jul 08 & 2829.35 & 20:26:38 & 320 & 6960 & 1.8569 & 3 & 10 \\
2003 Jul 09 & 2830.40 & 21:31:41 & 470 & 15891 & 2.0195 & 8 & 4 \\
2003 Jul 10 & 2831.37 & 20:59:29 & 420 & 14203 & 2.1715 & -2 & 5 \\
2003 Jul 11 & 2832.37 & 20:55:10 & 370 & 12054 & 2.3266 & 4 & 6 \\
2003 Jul 12 & 2833.35 & 20:29:36 & 410 & 13341 & 2.4793 & 18 & 5 \\
2003 Jul 16 & 2837.36 & 20:44:04 & 460 & 15960 & 3.1030 & 5 & 4 \\
2003 Jul 17 & 2838.36 & 20:41:13 & 350 & 12040 & 3.2582 & 8 & 6 \\
2003 Jul 18 & 2839.35 & 20:29:53 & 450 & 15685 & 3.4125 & 17 & 4 \\
2003 Jul 19 & 2840.35 & 20:25:25 & 450 & 15538 & 3.5675 & 10 & 4 \\
2003 Jul 20 & 2841.36 & 20:36:11 & 360 & 12187 & 3.7242 & 10 & 6\\
2003 Jul 21 & 2842.35 & 20:18:28 & 280 & 7407 & 3.8778 & 8 & 9 \\
2003 Jul 22 & 2843.35 & 20:23:19 & 440 & 14697 & 4.0339 & 10 & 5 \\
2003 Jul 24 & 2845.36 & 20:45:04 & 450 & 15398 & 4.3473 & 15 & 4 \\
2003 Jul 25 & 2846.36 & 20:37:52 & 480 & 16352 & 4.5020 & 12 & 4 \\
2003 Jul 27 & 2848.34 & 20:16:26 & 450 & 15747 & 4.8107 & 18 & 4 \\
2003 Jul 29 & 2850.34 & 20:15:25 & 510 & 16961 & 5.1217 & 5 & 4 \\
2003 Jul 30 & 2851.34 & 20:14:05 & 510 & 17672 & 5.2770 & 6 & 4 \\
2003 Jul 31 & 2852.34 & 20:07:15 & 540 & 18438 & 5.4318 & 16 & 4 \\
2003 Aug 02 & 2854.36 & 20:42:36 & 420 & 12907 & 5.7467 & 15 & 5 \\
2003 Aug 03 & 2855.34 & 20:12:51 & 420 & 13048 & 5.8990 & 3 & 6 \\
2003 Aug 04 & 2856.35 & 20:21:42 & 490 & 16195 & 6.0555 & 7 & 4 \\
\hline
\end{tabular}
\label{tab:journal}
\end{table*}

Observations were carried out at the T\'elescope Bernard Lyot, using the MuSiCoS \'echelle spectrograph (Baudrand \& B\"ohm 1992) associated to its Cassegrain-mounted polarimetric module (Donati et al. 1999). 

In polarimetric mode, MuSiCoS spectra cover the 450-660~nm range, with a resolving power close to 30,000. The analysis of circular polarization is performed with a quarter-wave plate (transforming circular polarization into linear polarization), followed by a Savart plate which splits the incident light into two beams, respectively containing light linearly polarized perpendicular/parallel to the axis of the beam-splitter. Each circularly polarized spectrum (also called Stokes V spectrum hereafter) is obtained from a series of four sub-exposures taken with the quarter-wave plate oriented at azimuth $\pm 45\degr$ with respect to the optical axis of the Savart plate. Following this procedure (detailed by Donati et al. 1999), both beams exchange their optical paths and positions on the detector, ensuring a removal of most spurious signatures to less than $5\times10^{-5}I_{\rm c}$ rms, where $I_{\rm c}$ denotes the continuum intensity. The intensity (Stokes I) spectrum is simultaneously produced by a simple addition of all four frames. Using a third combination of the series of sub-exposures, a ``null spectrum'' is obtained ($N$ hereafter, Donati et al. 1997), which does not contain any polarized signal.

The data set was collected during the summer of 2003, with 28 Stokes V spectra gathered from June 26 to August 04 (Table \ref{tab:journal}). The data reduction was performed using the ESpRIT package (Donati et al. 1997). In order to improve the wavelength calibration usually obtained for MuSiCoS spectra with reference to a Thorium-Argon spectrum, an additional calibration was performed, making use of atmospheric lines over-imposed on the stellar spectra. Using a line mask containing some 470 telluric lines, we thus reach a radial velocity accuracy of order of 300~m.s$^{-1~}$ (Petit et al. 2004a). The exposure time of each individual sub-exposure is 500~s, leading to a signal-to-noise ratio (\sn) of polarized spectra of the order of 550 in optimal weather conditions. In order to decrease further the noise level, the Least-Squares Deconvolution technique (LSD, Donati et al. 1997) was used to generate high \sn\ line profiles by simultaneously extracting the information from all photospheric lines available in the echellograms. Using a line mask calculated for a stellar atmosphere of spectral type G7, about 2,700 lines are available in the reduced spectra, allowing us to produce mean line profiles with \sn\ close to 20,000 ({\em i.e.} a typical \sn\ multiplex gain of 35 with respect to the peak \sn\ of the raw spectra). 

\begin{figure*}
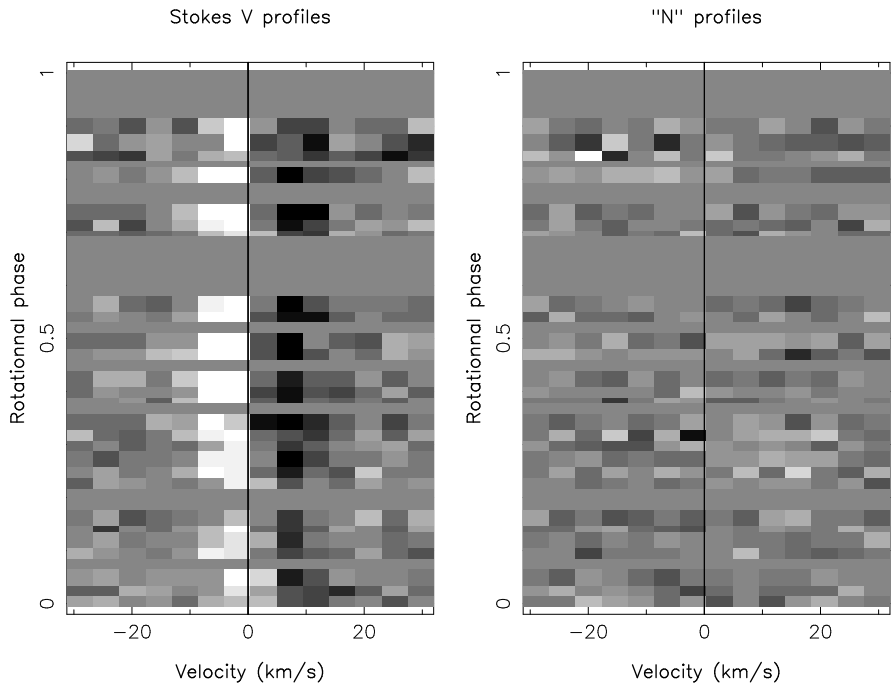

% *** Figure 1
\centerline{\mbox{\psfig{file=dynV.ps,height=9cm,angle=0}
\hspace{2mm}      \psfig{file=dynN.ps,height=9cm,angle=0}}}
\caption[]{Rotational modulation of line profiles. Left~: Stokes V profiles calculated with a G7 line-mask. Right~: ``Null'' profiles. White/black saturation levels correspond to $\pm3\times10^{-4}I_c$.}
\protect\label{fig:dyn}
\end{figure*}

\begin{figure*}
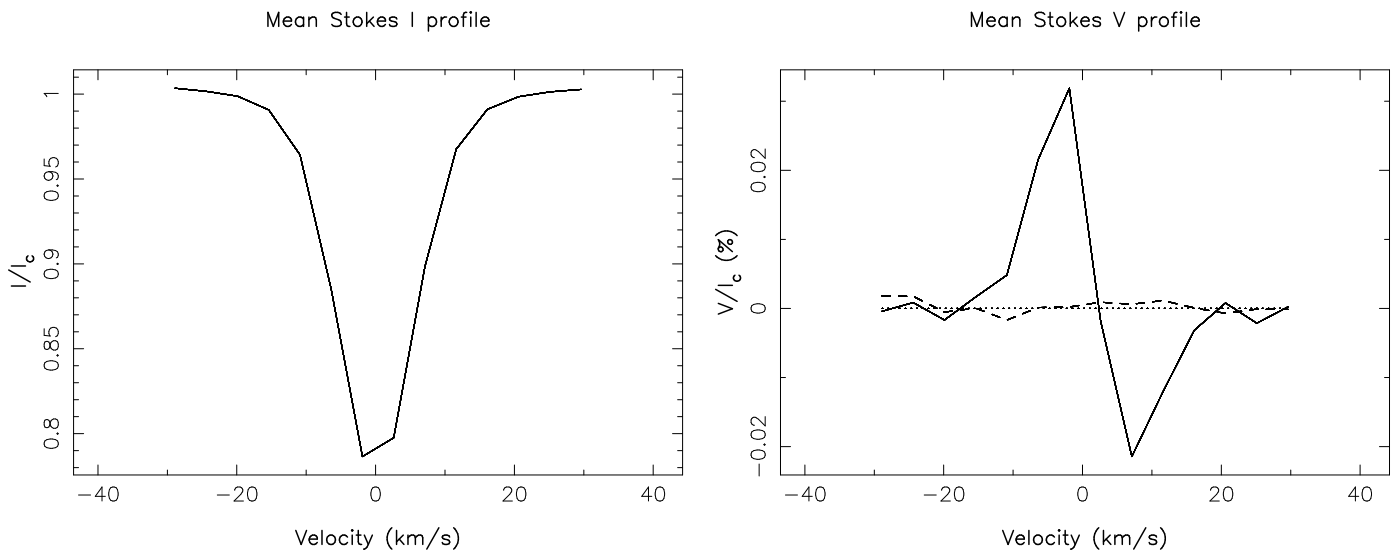

% *** Figure 2
\centerline{\mbox{\psfig{file=meanI.ps,width=9cm,angle=270}
\hspace{2mm}      \psfig{file=meanV.ps,width=9cm,angle=270}}}
\caption[]{Mean line profiles calculated from an average over all available observations. Left~: Stokes I profile. Right~: Stokes V profile (full line) and $N$ profile (dashes). The dotted line indicates the zero level.}
\protect\label{fig:mean}
\end{figure*}

The series of Stokes V profiles is plotted in Fig. \ref{fig:dyn}, in the rest-frame of the star, i.e. after correction of the heliocentric velocity (varying around an average value of -24~\kms) and of the mean stellar velocity (equal to 1.9~\kms\ at the date of observations). Zeeman signatures are repeatedly detected inside line profiles, always with a classical double-peaked shape appearing in the $\pm10$~\kms\ velocity range. The amplitude of signatures varies between 0.02\% and 0.07\% of the (non polarized) continuum level. The highest signal measured (on Jul. 18) has a significance of about 10$\sigma$, where $\sigma$ represents the uncertainty on the measurement. No spurious signature is detected in the $N$ profiles (Fig. \ref{fig:dyn}, right panel), confirming the robustness of the field measurements. We note also that spectra of magnetic and non-magnetic standard stars were obtained during the same observing run, verifying the nominal functioning of MuSiCoS. The global shape of Zeeman signatures is mostly stable with time (no polarity switch observed during the rotation, for instance), with a trend for their amplitudes to be larger around phase 0.5. 

A systematic asymmetry is observed between both lobes of the Stokes V profiles. If we call respectively $a_b$ and $a_r$ the amplitude of the blue and red lobe of the averaged profile (plotted in Fig. \ref{fig:mean}), we derive a relative asymmetry $\delta a = (a_b - a_r)/(a_b + a_r) = 0.19\pm0.05$. A difference is also visible in the areas of the lobes, with a relative asymmetry $\delta A = (A_b - A_r)/(A_b + A_r) = 0.22\pm0.04$ (where $A_b$ and $A_r$ are the areas of the blue and red lobe, respectively). No counterpart is seen in the mean $N$ profile, suggesting that such asymmetry is not of instrumental origin. We note that no obvious phase dependence of $\delta a$ and $\delta A$ is observed, owing to the limited \sn\ of the data set. 

The averaged Stokes I LSD profile is plotted in Fig. \ref{fig:mean}. The line asymmetry investigated by Toner \& Gray (1988) and Gray et al. (1996) using line bisectors is also observed in our data set, despite much lower spectral resolution. Bisectors are plotted in Fig. \ref{fig:bissectors}, for LSD profiles averaged over different bins in rotational phase. Each bin covers 20\% of the stellar rotation. We note that bisectors display a systematic shift toward the red near the continuum, which is consistent with previous findings of Toner \& Gray (1988). If we define the velocity span of a bisector as the difference in radial velocity between a point near the top of the profile ($I/I_c=0.95$) and a point close to the bottom ($I/I_c=0.8$), we find that the velocity span is marginally higher around phase 0.6 (with a span of about 170 \ms, vs. 110 \ms\ around phase 0.9). All other types of variability observed in Stokes I profiles are close in amplitude to the expected instrumental errors. 

\begin{figure}
% *** Figure 3
\centerline{\psfig{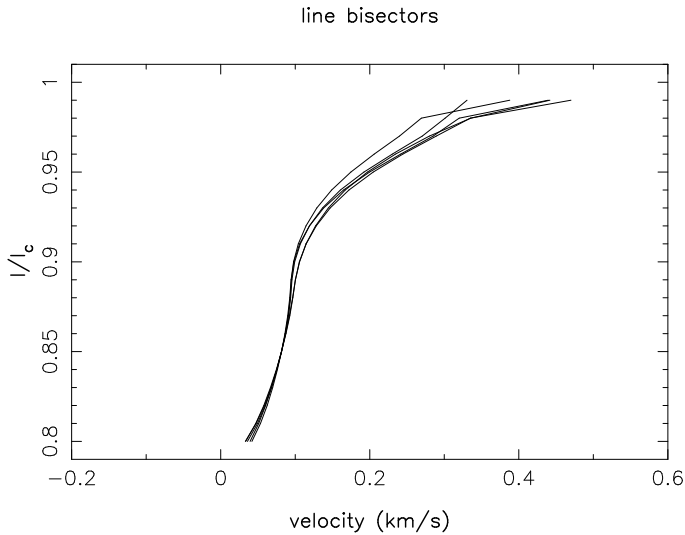}}
\caption[]{Bisectors of Stokes I LSD profiles averaged over different bins in rotational phase. Each bin covers 20\% of the rotation cycle, from phases 0.0-0.2 to phases 0.8-1.0.}
\protect\label{fig:bissectors}
\end{figure}

\section{Geometry of the large-scale field}
\label{sect:geometry}

\subsection{Longitudinal field measurements}
\label{sect:beff}

For every Stokes V and I profiles of the data set, we calculate the corresponding longitudinal magnetic field $B_l$ using the following expression (Donati et al. 1997, Wade et al. 2000)~:

\begin{equation}
B_l = -2.14\times10^{11}\frac{\int vV(v)dv}{\lambda_0 zc\int(I_c - I(v))dv}
\end{equation}

\noindent where $v$ is the radial velocity, $\lambda_0$ the mean wavelength of the line mask used to compute the LSD profiles (521~nm in our case) and $z$ the mean Land\'e factor of the line-list, equal to 1.21. The integration window covers a $\pm22$~\kms\ velocity range around the line centroid. The derived values of $B_l$ are listed in Table \ref{tab:journal}. The longitudinal field varies between -2~G and 18~G, with uncertainties ranging from 3 to 10~G. Significant fluctuations show up around the mean value of 8~G.

The $B_l$ time-series was used for a period search. The rotation period we obtain is $6.36\pm0.1$~d, in agreement with the period derived by Toner \& Gray (1988). The fact that no more than 6 consecutive periods are covered by our data-set explains the relatively large error bar we get here. 

\begin{figure}
% *** Figure 4
\centerline{\psfig{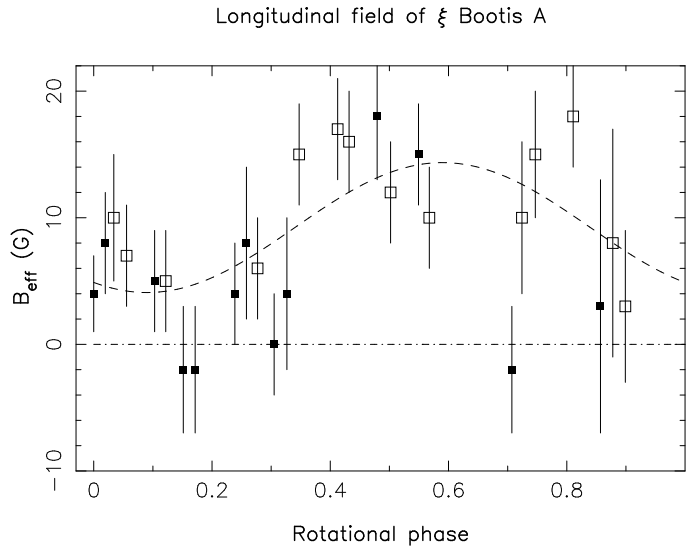}}
\caption[]{Longitudinal field as a function of the rotational phase for data collected until July 17 (filled squares) and after this date (open squares). The dashed curve represents a least-square fit of Preston's equation for the whole data set. The dash-dotted line illustrates a null magnetic field.}
\protect\label{fig:beff}
\end{figure}

We plot in Fig. \ref{fig:beff} the rotational modulation of $B_l$, according to the 6.43~d rotation period adopted in the paper. In the case of a tilted dipole, the rotational modulation of $B_l$ is described by the equation of Preston (1967)~:

\begin{equation}
B_l=B_{\rm p}\frac{15+u}{20(3-u)}(\cos \beta \cos i + \sin \beta \sin i \cos [2\pi (\phi -\phi_0)])
\end{equation}

\noindent where $B_{\rm p}$ is the field strength at the magnetic pole, $\phi_0$ the rotational phase at which the observer faces the positive magnetic pole and $\beta$ the inclination angle between the rotation and magnetic axis. $u$ is the limb-darkening coefficient, that we take equal to 0.66 according to Wade \& Rucinski (1985). The inclination $i$ is set to 28\degr. Making an iterative adjustment of Preston's relation, we obtain $B_{\rm p}=41\pm5$~G, $\beta=33\pm6$\degr and $\phi_0=0.59\pm0.04$, with a reduced \kis\ (\kisr\ from now on) equal to 1.2 (vs. \kisr=4.5 when assuming the field to be constant and equal to its mean value 8~G during the whole rotation). 

\subsection{Direct fitting of Stokes V profiles}

Rather than restricting the modeling of the field geometry to the first moment of Stokes V profiles, we conduct in this section a direct modeling of the profiles.

With this aim, we first check that the adopted values of stellar parameters enable a correct description of the Stokes I LSD profiles. The line model we adopt is similar to that described by Donati \& Brown (1997), except that the smearing induced by macro-turbulence is not assumed to be Gaussian in our case. A good fitting of stellar line profiles is obtained with \vsin=3~\kms\ and macro-turbulence $\zeta_{RT} = 5$~\kms, taken from Gray (1984). 

As a second step, we generate a synthetic time-series of Stokes V LSD profiles from an artificial star possessing a centered dipolar field, with rotational phases of the synthetic profiles taken equal to that of our observations. The model used to calculate polarized profiles is again that of Donati \& Brown (1997) where, for each pixel of the stellar surface, the local Stokes V profile is assumed to be proportional to the derivative of the local Stokes I profile. We then tune the values of $B_{\rm p}$, $\phi_0$ and $\beta$. Using a total of about 1,700 different values of the three field parameters, we generate as many synthetic data sets that we compare to the observations. A \kis\ is calculated for each model, so that we end up with a three-dimensional \kis\ landscape. The location of the minimum \kis\ gives the most likely value of the parameter triplet. 

Around the \kis\ minimum, error bars on the parameters (considered separately) correspond to an increase of \kis\ of unity (Press et al. 1992). Each error bar is multiplied by $(\chi_{\rm min}^2)^{1/2}$, in order to take partially into account potential systematic errors of the method (e.g. departure of the observed photospheric field from an inclined dipole). 

\subsection{Global dipole}
\label{sect:dipole}

\begin{figure*}
% *** Figure 5
\centerline{\psfig{file=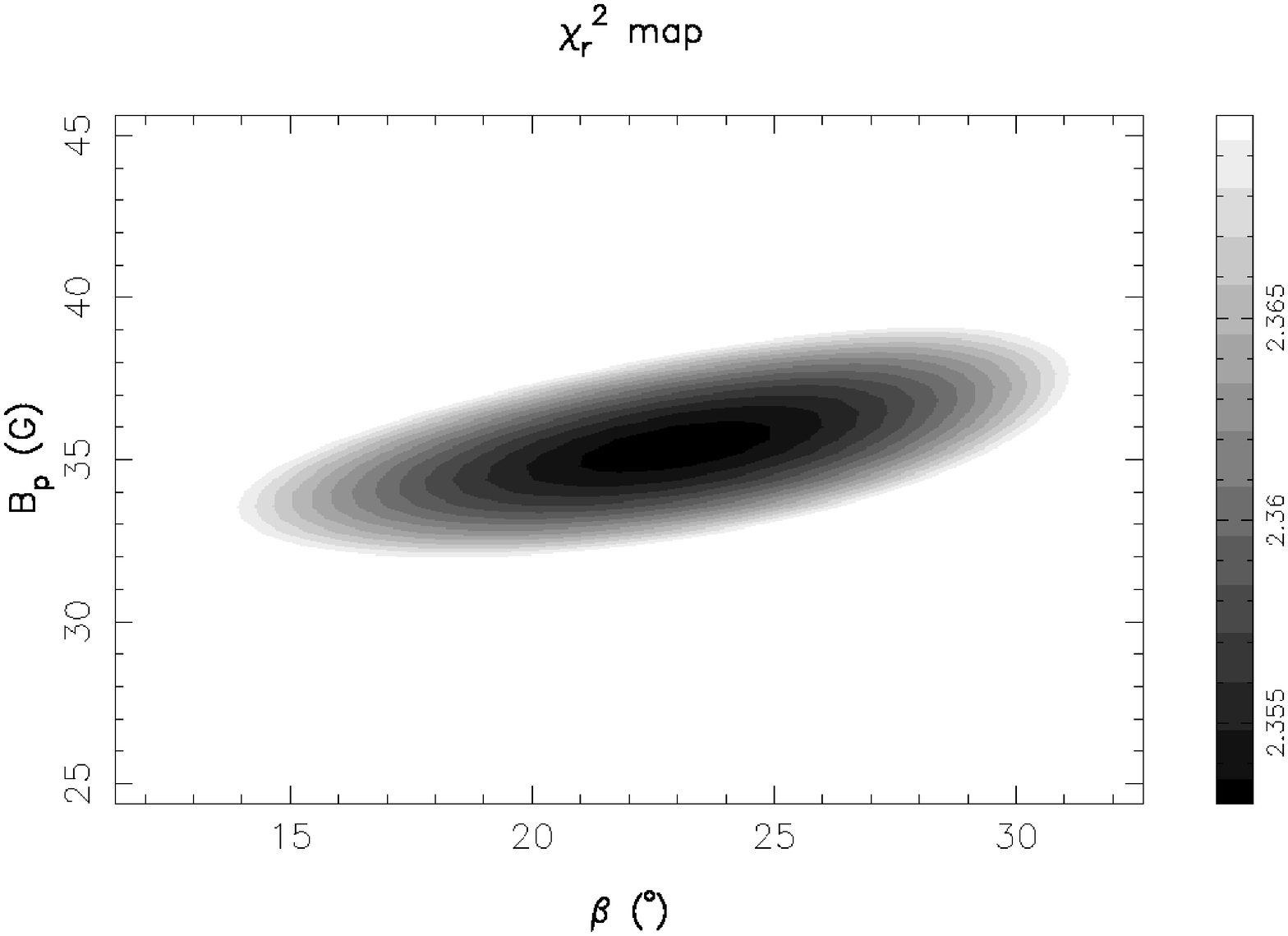,width=9cm,angle=270} \psfig{file=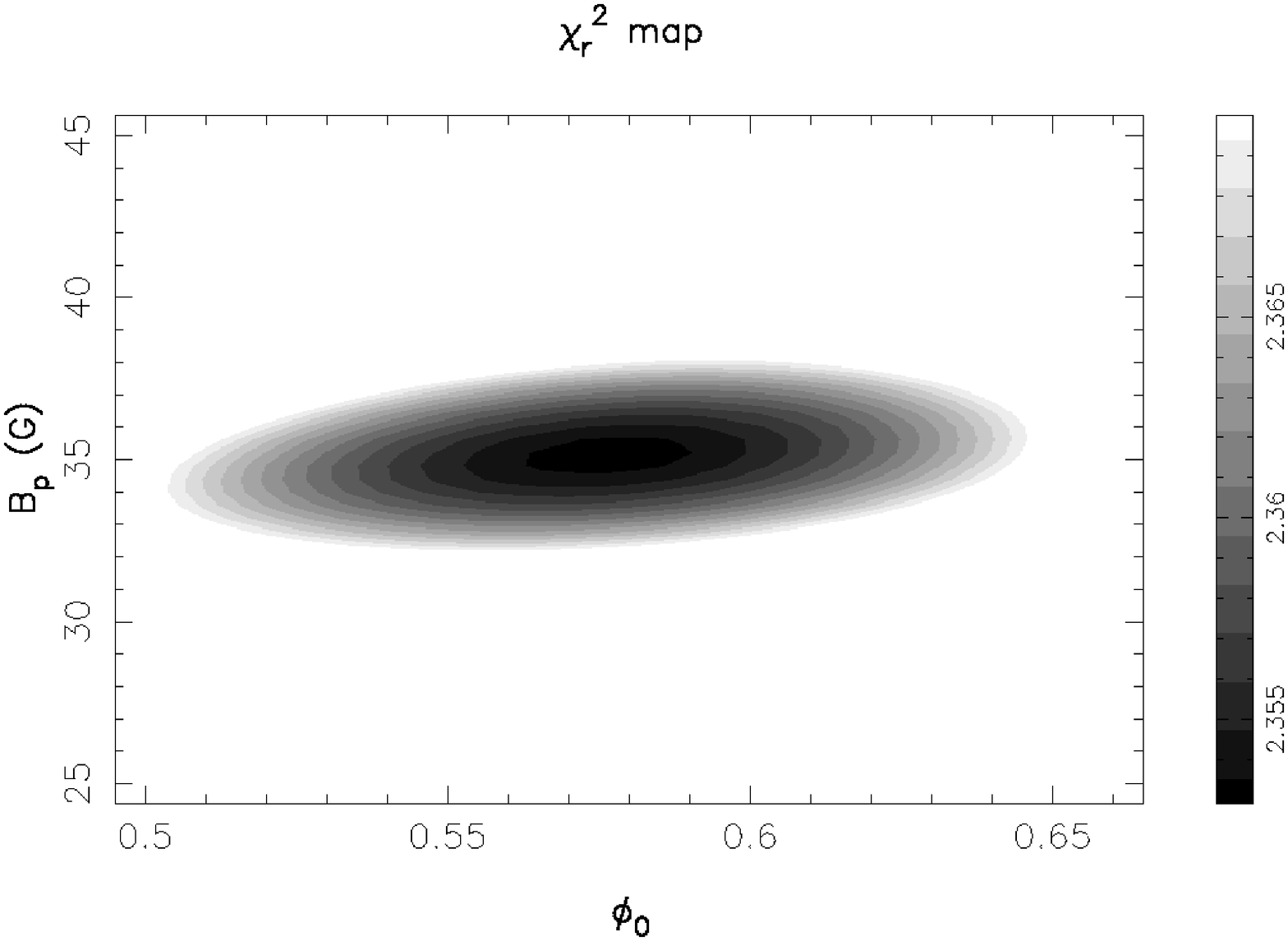,width=9cm,angle=270}}
\caption[]{\kisr\ maps obtained from slices in the three-dimensional field parameter space. $B_{\rm p}$ is the magnetic field strength at the magnetic pole. $\beta$ is the inclination of the dipole with respect to the rotation axis. $\phi_0$ is the rotational phase of maximum field strength.}
\protect\label{fig:dipolemaps}
\end{figure*}

Following this procedure, we derive $B_{\rm p} = 35\pm2$~G, $\phi_0=0.58\pm0.04$ and $\beta =22\pm5\degr$ (Tab. \ref{tab:param}). The minimum \kisr\ is not better than 2.3, suggesting that our model provides nothing more than a rough approximation of the profiles. In Figure \ref{fig:dipolemaps}, we represent 2D-maps that we extract from the 3D-\kisr-landscape by fixing one of the three free parameters to its best value. By doing so, we obtain a convenient visual control of the fitting procedure, showing that the \kisr\ minimum is unique, with well-defined ellipses of iso-\kisr\ surrounding the minimum. The synthetic profiles (averaged over the stellar rotation) are compared to the observations in Fig. \ref{fig:profilesdip}. 

As a second test of the robustness of our method, we run again the \kisr\ minimization, but this time using the $N$ profiles (following Wade et al. 2005). As expected, the derived field strength $B_{\rm p}$ is now compatible with zero  (Tab. \ref{tab:param}) and all values of $\beta$ and $\phi_0$ give similar \kisr\ for a fixed value of $B_{\rm p}$. Another major difference with the fitting of Stokes V profiles is that the minimum \kisr\ is now close to unity. The fact it is actually less than one (0.92 at minimum) suggests that error bars calculated for MuSiCoS spectra are over-estimated by about 4\% (as already pointed out by Wade et al. 2000). It tells also that the relatively high \kisr\ obtained with the Stokes V data set is probably due to intrinsic limitations in our magnetic model. 

At this stage, we note that the data fitting is not improved by going further in the multi-pole development of the field. If we add a quadrupolar field component (assumed to be aligned on the original dipole), the best fit is consistent with a null quadrupole (with a derived strength $B_{\rm q} = -5\pm5$~G). The \kisr\ is not significantly decreased by the addition of the quadrupole.

\begin{figure}
% *** Figure 6
\centerline{\psfig{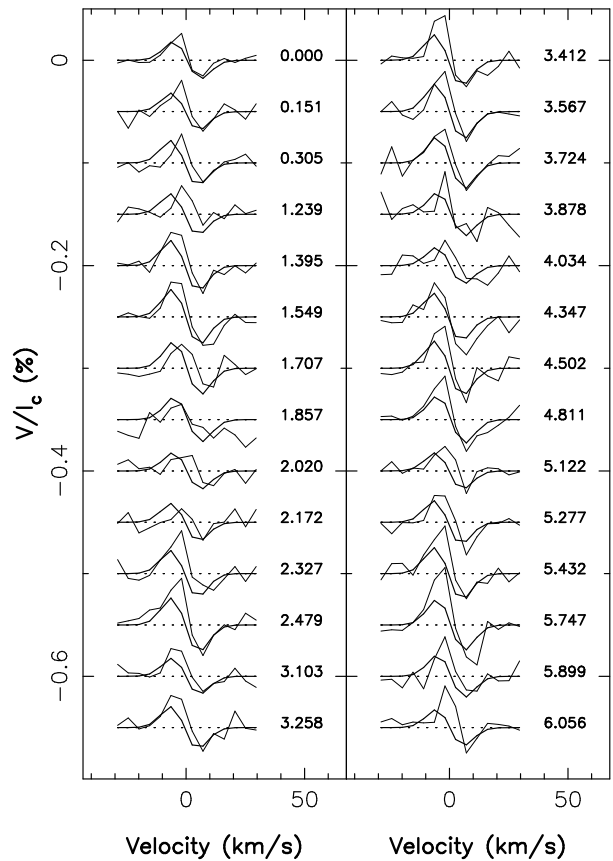}}
\caption[]{Synthetic Stokes V profiles (thick lines) compared to observed profiles (thin lines), in the case of a simple dipolar fit. Rotation cycles are indicated right to each profile. Successive profiles are shifted vertically, for display purpose. The dotted lines illustrate the zero level of each profile.}
\protect\label{fig:profilesdip}
\end{figure}

\subsection{Impact of stellar parameters}
\label{param}

In this section, we investigate whether a possible error in the determination of stellar parameters may contribute to the relatively high level of the \kisr\ we obtain with the fit of a dipolar field.

We first vary the inclination $i$. The main outcome of this test is to show that the magnetic model does not depend critically on this parameter. The field geometries we calculate for $i=20\degr$ and $i=40\degr$ are in good agreement with that we get for $i=28\degr$ (Tab. \ref{tab:param}). The best \kisr\ is still close to 2.3, with variations not exceeding 2\% for all three values of $i$.

A \kisr-landscape is also created for a rotation period of 6.14~d. Again, the \kisr\ is varying by less than about 1\% with respect to the original period of 6.43~d. New values of $B_{\rm p}$ and $\beta$ keep within the error bars of the original values. Because the rotational phases are calculated with the new value of the period, $\phi_0$ changes significantly and is now equal to $0.74\pm0.05$. 

From both tests, it is clear that the limited quality in data fitting we obtain with a purely dipolar field is unlikely to come from a wrong estimate of stellar parameters.

\section{Large-scale toroidal field}

Since a potential error in fundamental stellar parameters does not seem to be at the origin of the poor data adjustment, we propose in this section and in Sect. \ref{sect:granul} a series of tests, in order to understand better how various effects can lead to the observed limitations of the dipole model. Different options are therefore explored to improve the modeling of the magnetic geometry, and for each test we discuss the impact observed on the values of the field parameters, as well as on the value of the \kisr.

\begin{table*}
\caption[]{Estimated magnetic parameters for the various models described in the paper. Values without error bars are taken constant during the \kis\ minimization process. The best-fitting models are represented as bold rows.}
\begin{tabular}{cccccl}
\hline
$B_{\rm p}$     & $\phi_0$      & $\beta$  & $B_{\rm t}$       & \kisr\ & Notes\\
 (G)        &               & (\degr)  & (G)        &        & \\
\hline
 $35\pm2$   & $0.58\pm0.04$ & $22\pm5$ & 0          & 2.35   & dipole only\\
 $0\pm2$    & --            & --       & 0          & 0.92   & dipole, N profiles\\
 $33\pm2$   & $0.59\pm0.05$ & $22\pm7$ & 0          & 2.37   & dipole, inclination $i=20$\degr\\
 $39\pm2$   & $0.59\pm0.05$ & $18\pm7$ & 0          & 2.30   & dipole, inclination $i=40$\degr\\
 $34\pm2$   & $0.74\pm0.05$ & $21\pm7$ & 0          & 2.32   & dipole, rotation period of 6.14~d\\
 $33\pm2$   & $0.48\pm0.07$ & $23\pm7$ & 0          & 2.26   & dipole, odd-numbered profiles\\
 $31\pm1$   & $0.45\pm0.05$ & $22\pm6$ & 0          & 2.29   & dipole, even-numbered profiles\\
 $27\pm2$   & $0.64\pm0.13$ & $9\pm7$  & 0          & 1.92   & dipole, first half of the data set\\
 $40\pm2$   & $0.55\pm0.03$ & $28\pm4$ & 0          & 2.53   & dipole, second half of the data set\\
 ${\bf 39\pm2}$   & {\bf 0.58}          & ${\bf 35\pm3}$ & ${\bf -148\pm9}$ & {\bf 1.30}   & {\bf dipole + toroidal field}\\
 $0\pm2$    & 0.58          & --       & $0\pm8$    & 0.92   & dipole + toroidal field, N profiles\\ 
 $30\pm2$   & 0.58          & $24\pm4$ & $-124\pm8$ & 1.22   & dipole + toroidal field, first half\\ 
 $43\pm2$   & 0.58          & $35\pm3$ & $-160\pm8$ & 1.33   & dipole + toroidal field, second half\\ 
 $35\pm2$   & 0.58          & $31\pm4$ & $-137\pm8$ & 1.36   & dipole + toroidal field, odd-numbered prof.\\ 
 $37\pm2$   & 0.58          & $35\pm4$ & $-149\pm8$ & 1.32   & dipole + toroidal field, even-numbered prof.\\ 
 $37\pm2$   & $0.58\pm0.04$ & $23\pm5$ & 0          & 2.31   & dipole, 2-component atmosphere\\
 $37\pm2$   & $0.57\pm0.05$ & $22\pm6$ & 0          & 2.60   & dipole, rising flow only\\
 $401\pm26$ & $0.63\pm0.03$ & $36\pm4$ & 0          & 2.31   & dipole, sinking flow only\\
 $38\pm2$   & $0.58\pm0.03$ & $23\pm4$ & 0          & 1.87   & dipole, asymmetric synthetic profiles\\
 ${\bf 40\pm2}$   & {\bf 0.58}          & ${\bf 33\pm3}$       & ${\bf -112\pm8}$ & {\bf 1.23}   & {\bf dipole + toroidal field, asym. profiles}\\
\hline
\end{tabular}
\label{tab:param}
\end{table*}

\subsection{Large-scale toroidal field}
\label{sect:tor}

\begin{figure*}
% *** Figure 7
\centerline{\psfig{file=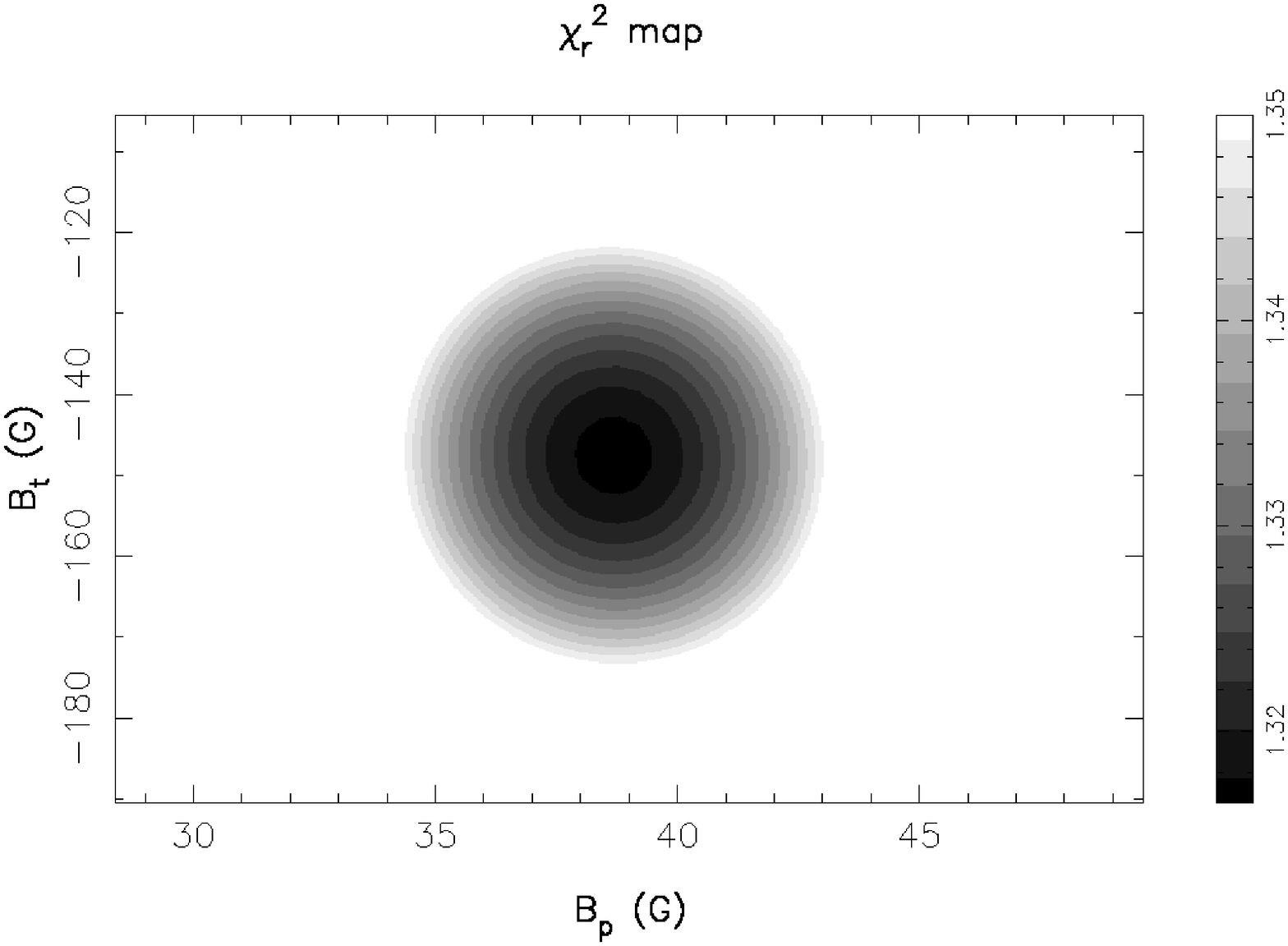,width=9cm,angle=270} \psfig{file=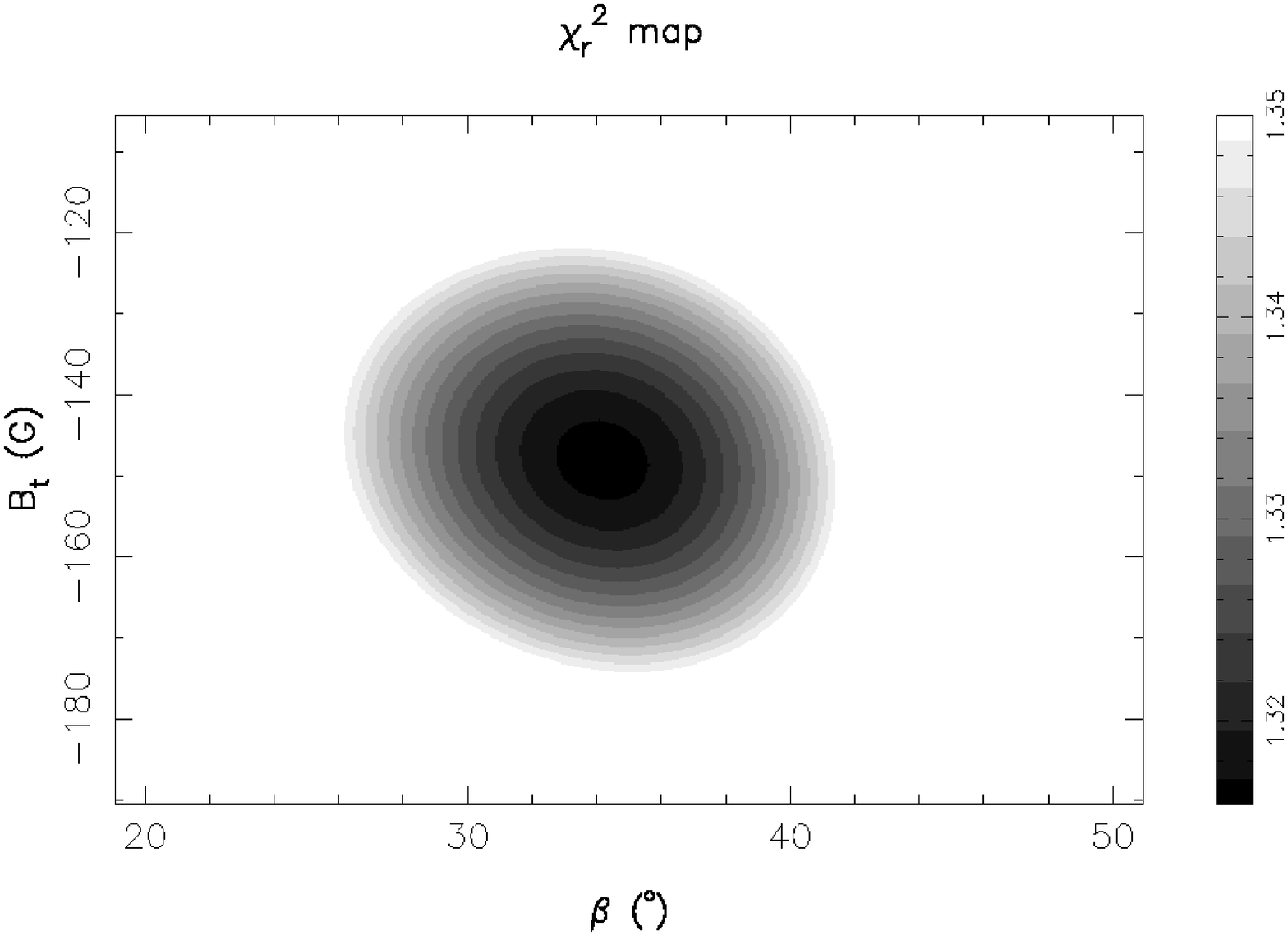,width=9cm,angle=270}}
\caption[]{\kisr maps obtained in the field parameter space. $B_{\rm p}$ is the magnetic field strength at the magnetic pole. $\beta$ is the inclination of the dipole with respect to the rotation axis and $B_{\rm t}$ the strength of the azimuthal component.}
\protect\label{fig:tormap}
\end{figure*}

Since a simple dipolar geometry of the surface field is not able to produce a satisfactory data fitting, we propose to modify our magnetic model by adding a new component to the large-scale photospheric field. We choose to impose the presence of an azimuthal field component (in addition to the inclined dipole). This addition is inspired by a result of ZDI, revealing the existence of large-scale azimuthal fields on most extremely active stars observed to date (e.g. Donati et al. 2003, Petit et al. 2004b).

As a very rough assumption, we take the strength of this component to be constant over a ring encircling the pole between latitudes 30\degr\ and 60\degr, which is somewhat reminiscent of the large-scale structure of this component on several rapidly-rotating stars observed so far. Because this ring possesses an axi-symmetric geometry, we assume that the determination of $\phi_0$ will not be affected by its presence. We therefore keep $\phi_0$ equal to 0.58 and adjust again three parameters simultaneously~: $B_{\rm p}$, $\beta$ and the toroidal field strength $B_{\rm t}$. 

The best fit is obtained for $B_{\rm t}=-148\pm9$~G (where $B_{\rm t}<0$ corresponds to a clockwise field). The value of $B_{\rm p}$ is slightly affected by the presence of the azimuthal field ($39\pm2$~G). $\beta$ is also slightly increased to $35\pm3$\degr. Slices taken from the 3D-\kisr-landscape are represented in Fig. \ref{fig:tormap}. We obtain here a very significant improvement in the quality of the fit, with \kisr\ now equal to 1.3. As previously done with the dipole, we check that the same \kis\ minimization, applied to the $N$ profiles, results in a toroidal component consistent with zero (Tab. \ref{tab:param}) and a \kisr\ close to unity. The value of the azimuthal field depends of course on the latitudinal extent of the ring on the synthetic star (mainly as a result of the different filling factor). Assuming for instance that the ring fills the whole hemisphere of the star reduces the field strength to $B_{\rm t}=-81\pm5$~G, but no significant impact is observed on the other field parameters, as well as on the minimum \kisr.

\subsection{Short-term evolution of the field geometry}
\label{subsets}

Despite very significant improvement with respect to a purely dipolar field model, a \kisr\ of unity is still not reached with the additional assumption of a toroidal field component. In this section, we investigate to what extent an intrinsic evolution of the field during data collection may produce such limitation.

In order to investigate potential short-term changes in the field geometry, we generate a couple of sub-sets from our original data set, the first one containing observations of the first half of the run (until July 17), while the second one contains observations gathered after this date. Both sub-sets present a correct phase sampling, as can be checked from Tab. \ref{tab:journal}. 

Assuming first a purely dipolar geometry, we determine the field parameters for each new data set. From the first one, we derive a field strength $B_{\rm p} = 27 \pm 2$~G and an inclination of the dipole axis $\beta = 9 \pm 7$\degr. The parameters derived from the second one are $B_{\rm p} = 40 \pm 2$~G and $\beta = 28 \pm 4$\degr. Both sub-sets give consistent values of $\phi_0$. The minimum \kisr\ is better for the first set (1.9 versus 2.5). We suggest that this apparent evolution is real and results from an intrinsic evolution of the stellar magnetic field over the 40 consecutive days of our observing window. A direct look at the $B_l$ measurements confirms this evolution. As an example, $B_l$ is equal to $B_l=18\pm4$~G and $15\pm5$~G respectively at rotation cycles 4.811 and 5.747 (July 27 and Aug. 02), while the longitudinal field does not exceed $-2\pm5$~G at phase 1.707 (July 07). We note that changes in the field geometry during data collection are not responsible for the high \kisr\ of the dipole model, since the \kisr\ level is not significantly decreased by the splitting of the original data set. 

As for the global dipole, we check if an evolution is also observed for the toroidal component. Using again the same couple of subsets, we calculate two \kisr-landscapes and optimize $B_{\rm p}$, $\beta$ and $B_{\rm t}$. We report an increase of $B_{\rm t}$, from $-124\pm8$~G to $-160\pm8$~G. The increase of $B_{\rm p}$ and $\beta$ is also observed here. Again, the splitting of the data set is not sufficient to provide a \kisr\ of unity (\kisr=1.22 and 1.33 for the first and second sub-sets, respectively). 

In order to double-check the reality of the field evolution, we now split the original data set into a new pair of sub-sets, the first one containing odd numbered profiles and the second one built up with even-numbered profiles. The dipole and toroidal field parameters obtained in both models are in close agreement with the values derived from the full data set (Tab. \ref{tab:param}). 

\section{Effect of surface velocity fields}
\label{sect:granul}

\begin{figure}
% *** Figure 8
\centerline{\psfig{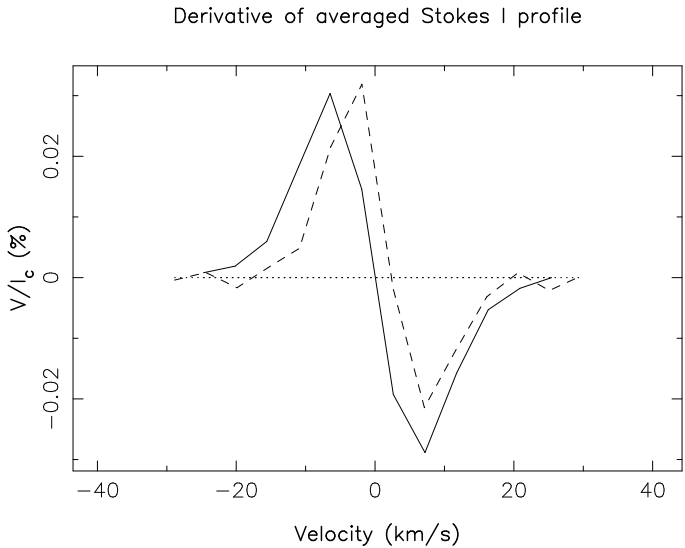}}
\caption[]{Derivative of the averaged Stokes I profile (full line), compared to the averaged Stokes V profile (dashes).}
\protect\label{fig:synstokes}
\end{figure}

In Fig. \ref{fig:synstokes}, we plot the derivative of the averaged Stokes I line profile, together with the mean Stokes V profile. The Zeeman signature presents a clear velocity shift with respect to the line centroid, showing up as a 2.5~\kms\ red-shift of the velocity at which the profile changes sign ( called zero-crossing velocity from now on). This shift is well above the 300~\ms\ radial velocity accuracy of the spectrograph. It reaches a maximum of about 5~\kms\ at rotation cycles 1.239, 1.707 and 6.056, but no evidence of a smooth phase dependence is observed. In this context, a simple magnetic dipole (producing Zeeman signatures roughly proportional to $dI/d\lambda$) cannot reproduce the observed shape of Stokes V profiles. A large-scale azimuthal field component mostly affects circularly polarized profiles when seen at intermediate limb angle. With an axi-symmetric ring of azimuthal field, two main regions of different radial velocities (on opposite sides of the meridian facing the observer) contribute to form two Zeeman signatures of opposite signs. The whole ring therefore produces a Zeeman signature resembling $d^2I/d\lambda^2$ and is able to compensate for the apparent velocity shift of Stokes V profiles.

In this section, we investigate to what extent a systematic velocity of magnetic regions (presumably related to surface convective flows) may also contribute to the observed red-shift of circularly polarized profiles.

\subsection{Two-component model of convective motions}

We adopt a modeling of stellar granulation inspired from the work of Gray \& Toner (1985). The surface convective flows are described in a two-component model, with hot (rising) and cool (sinking) material. Each component is assumed to produce a Gaussian line profile of dispersion $\zeta$, centered around radial velocity $v_{\rm cool}$ and $v_{\rm hot}$ for the down-flow and up-flow respectively. The line-depth ratio between both Gaussian components is quoted $R_{\rm c/h}$ thereafter. The granulation parameters we adopt are those Toner \& Gray (1988) derived specifically for \xib, with $\zeta=5$~\kms, $v_{\rm cool}-v_{\rm hot}=5.4$~\kms\ and $R_{\rm c/h}=0.1$. The adopted value $v_{\rm hot}=-0.4$~\kms\ is the only parameter we modify from the value proposed in the model of Toner \& Gray (in which $v_{\rm hot}=-1.4$~\kms). A slight red-shift of 1~\kms\ with respect to the original model is imposed, in order to take into account that the predominant hot component possesses a radial velocity close to zero in our data set (as a result of the correction of the mean radial velocity of the star). The same red-shift is imposed to $v_{\rm cool}$, in order to keep $v_{\rm cool}-v_{\rm hot}$ equal to the value of the original model (bearing in mind that this parameter, along with $R_{\rm c/h}$, controls the final shape of the line profile).

The main effect of this new line model is to introduce an asymmetry in the shape of Stokes I profiles. The level of asymmetry is however small and does not produce a significant impact in Stokes I fitting. For Stokes V, if we assume at first that the magnetic dipole is the same in granules and in inter-granular lanes ({\em i.e.} identical Zeeman signatures are formed in both components of the atmosphere), the data adjustment leads to a \kisr\ very similar to that obtained in Sect. \ref{sect:dipole} (see Tab. \ref{tab:param}). The best-fitting dipole parameters (Bp, $\beta$, $\phi_0$) remain also very close to the original values of Sect. \ref{sect:dipole}.

As a second step, we run again the \kis\ minimization procedure, but this time we keep only one of the two components of the convective atmosphere. By doing so, we assume that magnetic regions are concentrated either in the rising or sinking material alone. Considering first the rising component, the parameters of the dipole remain essentially unchanged. Using now the sinking component only, the derived field parameters are very different from anything we obtained before. While $\beta$ keeps its typical value ($36\pm4$\degr), $B_{\rm p}$ is now equal to $401\pm26$~G. This large increase in field strength is a direct consequence of the fact that the cool component of the atmosphere contributes for $1/10^{\rm th}$ only of the continuum flux, so that a magnetic field 10 times more intense than before is needed to account for the amplitude of Zeeman signatures. The minimum \kisr\ reached in both cases (2.6 and 2.3 for the hot and cool component respectively) shows that no improvement in data fitting is obtained.

As a last test, we calculate several 3D-\kisr-landscapes for different values of the vertical velocity of magnetic regions. The best result is obtained for a down flow of 2.3~\kms (in agreement with the observed red-shift of the zero-crossing). The dipole parameters are not very affected by this velocity shift, and the corresponding \kisr\ is equal to 1.4. We note that this velocity is quite far from the typical value of 5~\kms\ we would expect in the case of a convective down-flow. We explore further in the next section the possibility that magnetic elements may be confined in down-flows and show that this idea is not consistent with the basic properties of turbulent convection.

\subsection{Temperature of magnetic regions}
\label{bluered}

\begin{figure}
% *** Figure 9
\centerline{\psfig{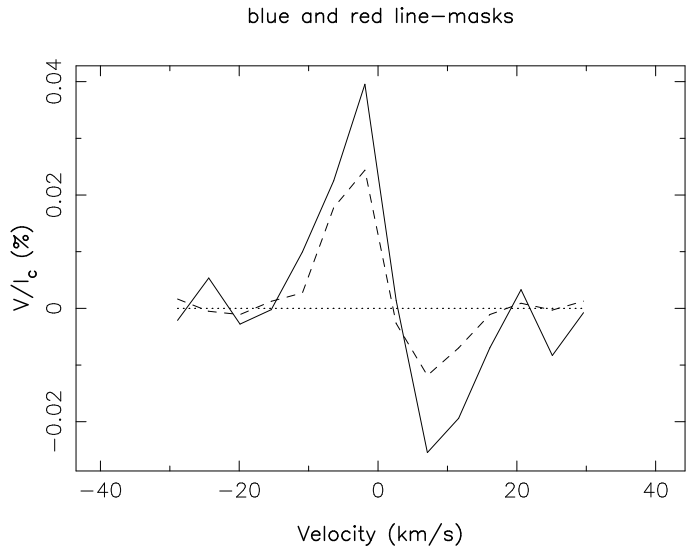}}
\caption[]{Averaged Stokes V profiles calculated with blue and red line-masks (full line and dashes, respectively).}
\protect\label{fig:bluered}
\end{figure}

If the apparent red-shift of polarized profiles is linked to a concentration of magnetic regions in down-flows, the classical picture of turbulent convection imposes that magnetic regions should be cooler than the quiet photosphere. To test this idea, we calculate two new sets of Stokes V LSD profiles, using two sub-masks built up respectively with the blue and red parts of the G7 line-mask. The wavelength limit is taken equal to 550~nm, ensuring that both sub-masks produce line-profiles of similar \sn. The mean wavelengths of the blue and red masks are 477~nm and 587~nm respectively. In both line-lists, the mean Land\'e factor is close to 1.21. In Fig. \ref{fig:bluered}, we illustrate the differences observed between the averaged blue and red Stokes V profiles. Note that the blue profile is slightly expanded to compensate for a difference in equivalent width of about 7\% between blue and red Stokes I LSD profiles.

A difference in the amplitude of Zeeman signatures is visible, with peaks at $2.4\times10^{-4}\pm2\times10^{-5}I_{\rm c}$ and $3.9\times10^{-4}\pm2\times10^{-5}I_{\rm c}$ for the red and blue signatures, respectively. If the Zeeman signatures are formed uniformly in the photosphere (of temperature 5550~K), no significant difference should be observed between the blue and red profiles. The blue excess we can see here therefore suggests that Zeeman signatures are preferentially formed in hotter regions of the photosphere. If we exclusively link the blue excess to a temperature effect, the implication is that magnetic regions are several thousand Kelvin hotter than the quiet photosphere. This temperature difference is too important to be considered as a satisfactory explanation. We therefore suggest that additional (still not understood) effects must be taken into account to reproduce the observed blue/red contrast of Stokes V profiles.

If magnetic elements producing the observed Zeeman signatures are hotter than the average temperature of the photosphere, no red-shift of the signatures should be observed, at least as long as turbulent convection is concerned. We therefore conclude that the apparent red-shift of Stokes V profiles is not related to a systematic vertical velocity of magnetic regions. In this context, the presence of a large-scale toroidal field provides a much more satisfactory interpretation of the observed shape of polarized profiles.

\subsection{Asymmetry of Stokes V profiles}
\label{sect:asymmetry}

\begin{figure}
% *** Figure 10
\centerline{\psfig{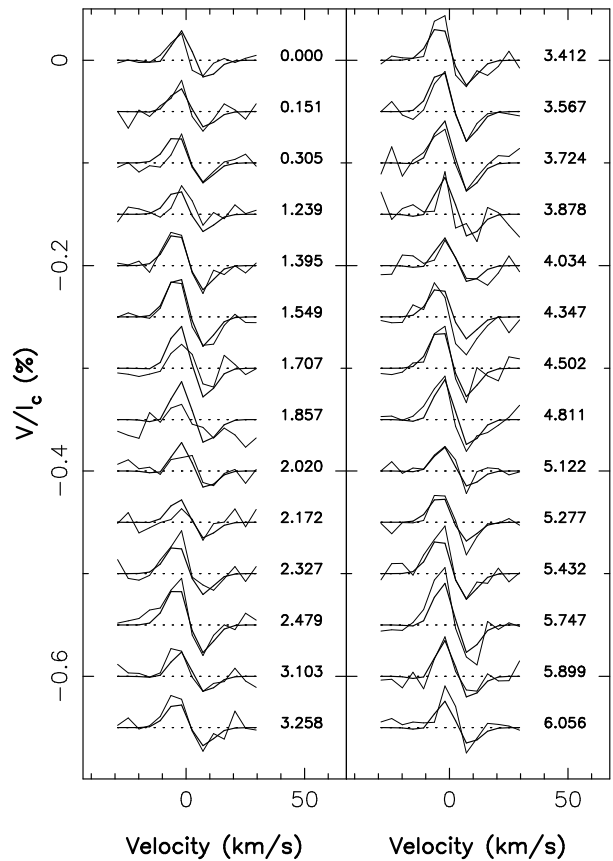}}
\caption[]{Same as Fig. \ref{fig:profilesdip}, for a magnetic model including an inclined dipole, a toroidal field and an asymmetry of Stokes V profiles.}
\protect\label{fig:profilestorasym}
\end{figure}

\begin{figure*}
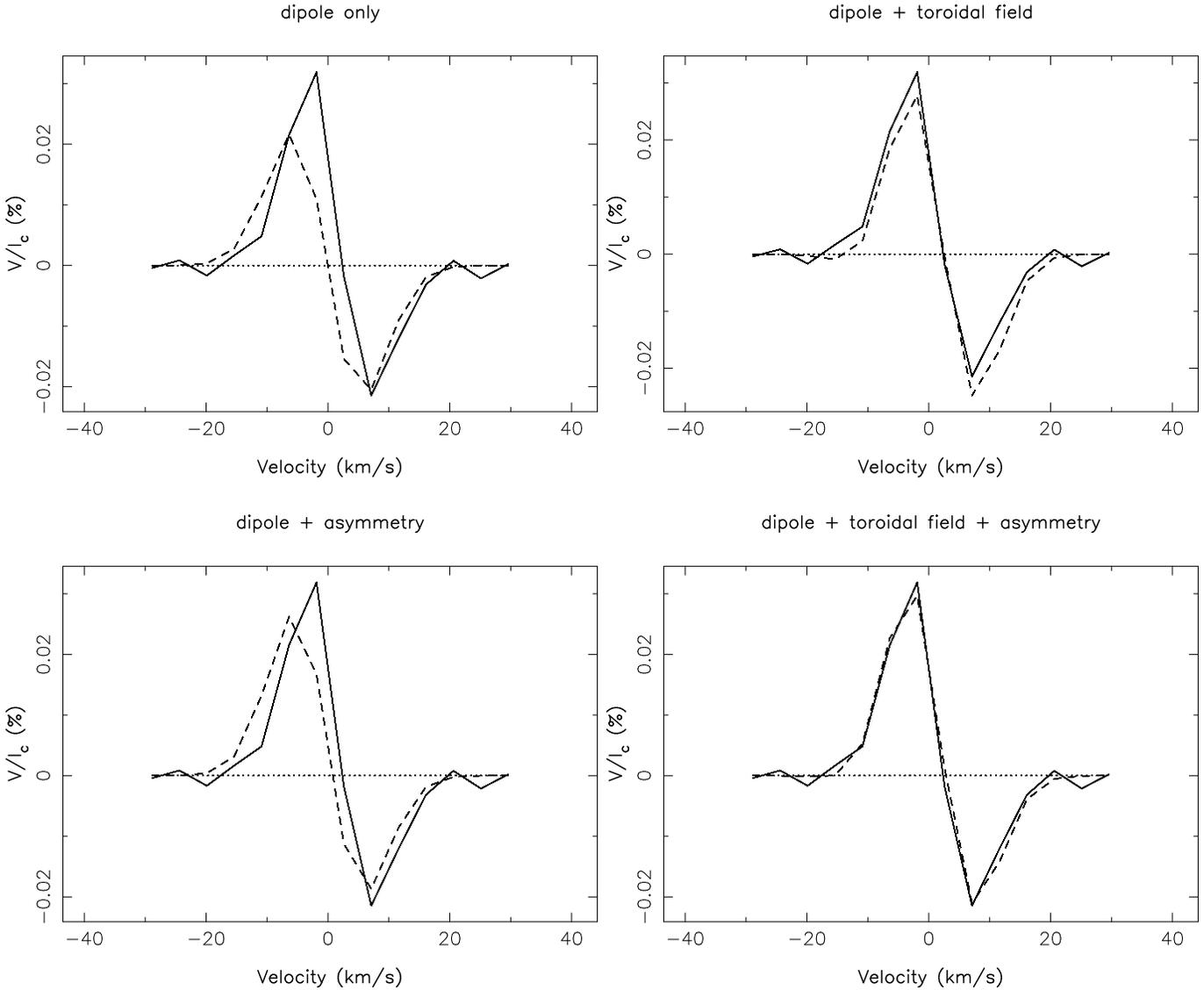

% *** Figure 11
\centerline{\psfig{file=meanbestdip.ps,width=9cm,angle=270} \psfig{file=meanbesttor.ps,width=9cm,angle=270}}
\vspace{5mm}\centerline{\psfig{file=meanbestdipassym.ps,width=9cm,angle=270} \psfig{file=meanbesttorassym.ps,width=9cm,angle=270}}
\caption[]{Averaged observed Stokes V profile (full line) compared with mean profiles obtained from different modeling options described in the text (dashes). Left panels~: dipolar field without and with asymmetric synthetic profiles (top and bottom, respectively). Right panels~: dipole and azimuthal field without and with asymmetry (top and bottom, respectively).}
\protect\label{fig:asymmetry}
\end{figure*}

As already pointed out in Sect. \ref{sect:obs}, an asymmetry is observed in Stokes V profiles, with an amplitude and area of the blue lobe exceeding that of the red lobe. This asymmetry can also be readily seen in Fig. \ref{fig:synstokes}, where the (symmetric) derivative of Stokes I profile is compared to the averaged Stokes V profile. We note that the level of asymmetry is higher for the mean profile calculated with a red line-masks (Fig. \ref{fig:bluered}). For the blue profile, we derive $\delta A = 0.17 \pm 0.06$ and $\delta a = 0.22 \pm 0.07$. For the red profile, $\delta A = 0.34 \pm 0.08$ and $\delta a = 0.35 \pm 0.09$ (where $\delta A$ and $\delta A$, as defined in Sect. \ref{sect:obs}, represent the relative asymmetry affecting respectively the area and amplitude of the lobes). 

In this section, we introduce an asymmetry in our synthetic polarized profiles and test whether this new parameter can affect the reconstructed field geometry. We simulate asymmetric local Stokes V profiles by expanding one lobe of the profile and reducing the other lobe by the same factor. The local profile is then smeared by the instrumental resolution and by the rotational broadening. The effect of this smearing is to decrease the amplitude of the profiles and generate an apparent shift of the zero-crossing wavelength (as described by Solanki \& Stenflo 1986). The best fit of the mean Stokes V profile is obtained for a relative asymmetry of the {\em local} profile $\delta a \approx \delta A \approx 0.11$, which is then used to calculate magnetic models.

Assuming a dipolar field geometry as in Sect. \ref{sect:dipole}, the addition of the asymmetry factor does not lead to any significant impact on the dipole parameters. The \kisr\ is however significantly improved and is now equal to 1.87 (vs. 2.35 without asymmetry). However, this relatively high \kisr\ value demonstrates that an asymmetry factor is not sufficient to ``mimic'' the shape of observed Stokes V profiles (Fig. \ref{fig:asymmetry}).

Assuming a field geometry composed from a global dipole and an axi-symmetric toroidal field, the value of $B_{\rm p}$ keeps close to that obtained with symmetric profiles. By contrast, the value of $B_{\rm t}$ is significantly decreased, from $-148\pm9$~G to $-112\pm8$~G. This evolution is due to the red-shift of the zero-crossing velocity induced by the combination of a local asymmetry and line smearing. Here the \kisr\ is also slightly improved with respect to the original model, with a value of 1.23. The fit of individual phases is shown in Fig. \ref{fig:profilestorasym}.

\section{discussion}

\subsection{Large-scale dipole}

The modeling of our data set suggests that one component of the large-scale magnetic field of \xib\ is an inclined dipole, with a strength of about 40~G and an inclination angle of order of 35\degr\ with respect to the rotation axis. The dipole parameters are only slightly varying for different basic assumptions in our magnetic models (Tab. \ref{tab:param}).

Using observations collected in 1998, along with archival magnetic field measurements by Borra et al. (1984) and Hubrig et al. (1994), Plachinda \& Tarasova (2000) report a sign reversal of the longitudinal field $B_l$ during the stellar rotation, which denotes a relatively high inclination of the dipole with respect to the rotation axis ($\beta = 87\degr$ for their chosen inclination $i=40\degr$). It is obvious from Fig. \ref{fig:dyn} that this behavior is not observed in our data set. A quick look at our series of profiles is sufficient to conclude that the sign of Zeeman signatures is constant over the whole rotation, which translates into a constant sign of $B_l$ (Fig. \ref{fig:beff}). Our own estimate of $\beta$ ranges between 20\degr\ and 40\degr, depending on the details of the underlying magnetic model. This estimate is not very sensitive to $i$ and to the rotation period (Sect. \ref{sect:param}), which are different in our study from the values taken by Plachinda \& Tarasova. We note that, among all measurements used by Plachinda \& Tarasova, only six possess a negative value in excess of 1$\sigma$ (and only one in excess of 2$\sigma$). Five of these points come from observations gathered at the Crimean Astrophysical Observatory in 1998, and only one from archival measurements. One possibility is that negative $B_l$ values reported by Plachinda \& Tarasova come from intrinsic biases in their measurements. A second possibility is that the field geometry has evolved between their observations and our own observing campaign. Long-term changes in the photospheric field would also help to interpret the apparent discrepancy between their maximum $B_l$ values (about 50 to 80~G) and those reported in the present article (never exceeding 20~G). We come back to this last point in Sect. \ref{sect:evol}

\subsection{Large-scale toroidal field}

A simple dipole is not able to provide a good fit to our time-series of Stokes V profiles. This situation is dramatically improved by the addition of an axi-symmetric ring of azimuthal field on the visible hemisphere of the star. A large-scale toroidal field is also required if we take into account the profile asymmetry in our magnetic model, but in this case the intensity of the azimuthal component is decreased by about 20\%. 

If we do not introduce any azimuthal field component, the only option to obtain a similar \kis\ level is to assume that magnetic regions are concentrated in photospheric down-flows, with a vertical velocity of order of 2.3~\kms. We consider that this model is not adapted, for two main reasons. The first argument is that convective down-flows would correspond to material cooler than the quiet photosphere, while we deduce from the blue/red contrast of Zeeman signatures that the magnetic regions are, to the opposite, hotter than the average photospheric temperature. The second argument is that typical velocities of magnetic elements observed on the solar surface are even smaller than typical solar convective velocities and generally do not exceed a few hundreds of \ms (Solanki 1993), which is far below the value required here. Since convective velocities increase with surface temperature (Gray \& Toner 1985), we even expect potential systematic velocities of magnetic elements to be smaller on \xib\ than on the Sun, and in any case much smaller than 2.3~\kms. From these two arguments, we conclude that a toroidal field on \xib\ is the only acceptable interpretation of our spectropolarimetric observations.

Azimuthal magnetic fields were previously detected on a small number of fast-rotating stars (e.g. Donati et al. 2003, Petit et al. 2004b) in striking contrast to the solar case, where the largest magnetic structures generally host a vertical field (except in the penumbra of sun-spots). In a few cases, the toroidal component reconstructed by ZDI is sufficiently structured to display a clear latitudinal dependence, such as rings encircling the pole at different latitudes.

The possible detection of a toroidal field on \xib, obtained with a modeling technique independent from ZDI, suggests that large-scale azimuthal fields may be common on very active solar-type stars. Despite its low projected rotational velocity, the rotation period of \xib\ is equal to 6.43~d, which makes it a much faster rotator than the Sun. Since it has been known for a very long time that the efficiency of dynamo processes is correlated to the stellar rotation rate (e.g. Baliunas et al. 1995), it is not a surprise that we observe here magnetic characteristics already reported for intermediate rotators of rotation periods of a few days.

\subsection{Temporal evolution of the magnetic field}
\label{sect:evol}

We report a significant short-term evolution of the field geometry during the 40 consecutive nights of data collection. This evolution translates into an increase of $B_{\rm p}$, $\beta$ and $B_{\rm t}$. At this stage, it is not possible to decide if this changing geometry is due to a global evolution of the large-scale field, or to the continuous redistribution of active regions. Considering the rather short timescale of the evolution, we suggest the second option is more likely. Such fluctuations in the parameters of the large-scale field are expected, since the ``global field'' we measure here is the average, over the visible hemisphere of the star, of a potentially very complex and fluctuating distribution of individual magnetic regions. Future observations may tell us however if we observed here the first steps of a more global geometry change.

One way to interpret the apparent discrepancy between the magnetic geometries derived in the present study and in the previous work of Plachinda \& Tarasova (2000) is to invoke an intrinsic evolution of the global field. This would imply that the field geometry has significantly evolved in less than 5~yr (since 1998), with a marked decrease of the dipole inclination (from 87\degr\ to about 30\degr), together with a decrease in the dipole strength (from about 80~G to 40~G). This behavior would be very different from what is observed on the Sun, where the large-scale dipole is almost aligned on the rotation axis (at least when its strength is maximum, {\em i.e.} during solar minimum). Recent numerical simulations of magnetic flux transport processes on the solar surface suggest that the polarity reversal of the global dipole can be interpreted as a progressive 180\degr\ shift of its inclination (Baumann 2005). However, this short transition is occurring at solar maximum, when the strength of the dipole is very small (to a level where the dipole is probably not detectable). For \xib, this general picture is in contradiction with the apparent long-term evolution of the field geometry (suggesting that a higher inclination is associated to a stronger dipole). Only a monitoring of \xib\ over several years with a unique instrument, together with a field modeling performed with a unique procedure, will make it possible to determine if such non-solar field evolution is genuine. In the meantime, we cannot exclude that the observed differences with previous field measurements by Plachinda \& Tarasova partly arise from differences in instrumental configuration and/or data processing.

\subsection{Nature of magnetic regions}

The large-scale field investigated in the present study results from an average, over the visible hemisphere of the star, of magnetic structures of possibly various natures. The fact we observe a short-term evolution of the field suggests that individual magnetic regions at the origin of the large-scale field have typical life-times of days to weeks at most. Moreover, the blue/red contrast of Stokes V profiles indicates that magnetic elements we observe may be hotter than non-magnetic regions of the stellar surface. Both observations can be reconciled with what is observed on the Sun in faculae and in the network field. However, rotational variations in bisectors of Stokes I profiles provide a marginal evidence that the granulation velocity dispersion is higher for rotation phases at which the magnetic field is maximum. If confirmed by future observations (benefiting from a better spectral resolution), this may imply that the general characteristics of magnetic regions are very different from anything observed on the Sun. In particular, this could suggest that the ``star-patch'' reported by Toner \& Gray (1988) is associated to a large-scale structure of the photospheric field.

In order to reproduce the observed asymmetry of Stokes V profiles, we suggest that the {\em local} polarized profiles (i.e. prior to instrumental and rotational smearing) must possess a relative asymmetry $\delta a \approx \delta A \approx 0.11$. The value of $\delta a$ is similar to what is observed on the Sun (e.g. Pantellini et al. 1988). $\delta A$ is also close to solar measurements, though slightly higher than typical solar values. The asymmetry of solar profiles is usually modeled by the combined use of velocity and magnetic gradients in magnetic elements (see Solanki 1993 for a review). The sign and level of $\delta a$ and $\delta A$ depend on the nature of magnetic regions in which the Zeeman signature is formed (faculae and network field are considered by Solanki 1989, sunspot penumbra by Solanki \& Montavon 1993). In the future, a rigorous modeling of the asymmetric profiles of \xib\ may help to distinguish between the different possible magnetic structures at the origin of the observed asymmetry.

\section{Concluding remarks and future work}

Our direct modeling of the circularly polarized line profiles of \xib\ suggests that the large-scale field is mainly constituted of two main components, with a global dipole and a large-scale toroidal field. Using different magnetic models, we derive a dipole strength varying between 27 and 43~G, a dipole inclination ranging from 9\degr\ to 35\degr\ and a toroidal field strength spreading between 112 and 160~G. The observed range in magnetic parameters have several identified origins. First, the dipole parameters are affected by the addition of the toroidal field in our models. Second, the toroidal field strength itself is modified by the further assumption that Stokes V profiles are asymmetric. Third, all field parameters (dipole and toroidal component) are affected by a short-term evolution of the field geometry over the 40~nights of our observing window (with an increase of field strength and dipole inclination).

This work confirms that azimuthal field components may be common on very active solar-type stars (see also results derived from ZDI~: Donati et al. 2003, Petit et al. 2004b). Such information cannot be extracted from modeling techniques based on $B_l$ measurements, which make use of a small fraction only of the information available in the data. We demonstrated in the present study that, even for slow rotators with narrow line-profiles, tight constraints on the magnetic geometry can be obtained with a modeling of the full profiles. The detection of a Stokes V asymmetry is another example of information that can only be obtained using high \sn\ polarized profiles.  

We have engaged a long-term monitoring of \xib, in order to investigate the secular evolution of its global field. We will test in particular whether the large-scale field undergoes periodic polarity reversals (as in the solar case) or if fluctuations of the magnetic geometry are more erratic (as suggested by Baliunas et al. 1995 from a long-term monitoring of chromospheric activity). We will also follow potential long-term fluctuations in the blue/red contrast of Zeeman signatures and in the line asymmetry.

This work opens exciting perspectives for the study of slowly-rotating solar-type stars with moderate to low magnetic activity, for which classical Doppler imaging techniques are not applicable. We have demonstrated that their global field can be investigated with an accuracy of a few Gauss, and we expect even better performances with the new generation of spectropolarimeters like ESPaDOnS/CFHT and NARVAL/TBL, which will cover a broader spectral domain and benefit from a higher spectral resolution (Donati et al. 2005). The efficiency of multi-line techniques will be significantly increased with such instruments and therefore enable the investigation of dynamo cycles on a sample of solar-type stars, with the aim to estimate how various stellar parameters may impact general properties of the large-scale dynamo at work in active stars. 

\section*{ACKNOWLEDGMENTS}

PP thanks I. Baumann, A. Collier Cameron, M. Sch\"ussler and S.K. Solanki for fruitful discussions about the content of this paper. GAW and JDL acknowledge grant support from the Natural Sciences and Engineering Research Council of Canada (NSERC). We are grateful to an anonymous referee, whose comments helped to clarify the article.

\end{document}